\documentclass{article}

\usepackage{arxiv}

\usepackage[utf8]{inputenc} 
\usepackage[T1]{fontenc}    
\usepackage{hyperref}       
\usepackage{url}            
\usepackage{booktabs}       
\usepackage{amsfonts}       
\usepackage{nicefrac}       
\usepackage{microtype}      
\usepackage{lipsum}		
\usepackage{graphicx}
\usepackage{gensymb}
\usepackage{doi}
\usepackage[numbers]{natbib}

\title{A Simple and Novel Passive Double-Sensitivity Optical Gyroscope Based on Non-Reciprocal Polarization Techniques}

\author{
  Onder Akcaalan\\
  Independent Researcher \\
  Hamburg, Germany\\
  \texttt{\ onderakcaalan@gmail.com*} \\
     \And
 Melike Gumus Akcaalan\\
  Institute for Nanostructure and Solid-State Physics\\
  University of Hamburg \\
  Hamburg, Germany\\
  \texttt{\ melike.gumus.akcaalan@uni-hamburg.de} \\
}

\renewcommand{\shorttitle}{\textit{arXiv} Template}

\hypersetup{
pdftitle={A template for the arxiv style},
pdfsubject={q-bio.NC, q-bio.QM},
pdfauthor={David S.~Hippocampus, Elias D.~Striatum},
pdfkeywords={First keyword, Second keyword, More},
}

\begin{document}
\maketitle

\begin{abstract}

This paper presents a novel interferometric fiber optic gyroscope (IFOG) architecture, the Double-Sensitive Non-Reciprocal Polarization Phase Shifter IFOG (DS-NRPPS-IFOG), which introduces—for the first time—a fully passive phase biasing scheme capable of simultaneous operation at two quadrature points: $\pi/2$ and $3\pi/2$. Building upon prior passive biasing techniques, this design uses a Non-Reciprocal Polarization-Dependent Phase Shifter (NRPPS) combined with a double-pass sensing coil arrangement to achieve both passive $\pi/2$ phase modulation and enhanced measurement sensitivity. The system utilizes polarization manipulation and a quarter-wave retarder to create a double-sensitive response while eliminating the need for active modulators. Simulation results demonstrate significant performance improvements, with Angular Random Walk (ARW) values up to 50× lower than those of conventional DS-IFOG systems, depending on fiber length. Moreover, the architecture enables continuous rotation measurements and offers spontaneous noise suppression by leveraging dual quadrature detection. These findings mark a major advancement toward low-power, highly stable, and compact passive optical gyroscopes for precision navigation applications.

\end{abstract}

\keywords{IFOG \and gyroscope \and passive \and double \and sensitivity \and noise \and bias \and non-reciprocity}

\section{Introduction}

Interferometric Fiber Optic Gyroscopes (IFOGs) play an important role in a wide variety of precision navigation and guidance systems, including aerospace, maritime, defense and autonomous vehicles in both commercial and military applications. They are highly valued for their exceptional sensitivity \cite{sanders1996fiber} reliability \cite{lefevre2013fiber}, and long-term stability \cite{korkishko2012interferometric}. As solid-state devices with no moving parts \cite{grattan2000fiber}, IFOGs offer superior resistance to mechanical shock and vibration, making them ideal for use in harsh or dynamic environments. They rely on the Sagnac effect \cite{sagnac1913ether}, where light splits into clockwise (CW) and counter-clockwise (CCW) beams in a fiber coil. Rotation causes a phase difference proportional to angular velocity, which is detected through interference at a photodetector. A fixed phase bias, typically $\pi/2$, is introduced between the beams to place the system in the linear response region for optimal sensitivity. Traditionally, this bias is achieved using active phase modulators like electro-optic or piezoelectric devices \cite{lefevre2022fiber}, but these require complex control electronics, dedicated power, and thermal stabilization \cite{mao2005research,wang2014compensation,kiraci2017temperature}, adding to system size, cost, and susceptibility to aging, drift, and mechanical failure \cite{bi2018influence,wang2010study,sun2010study}. For high-sensitivity applications with longer coils, thermal phase noise (TPN) remains a key limiting factor \cite{takei2023simultaneous,song2017modeling}, and piezoelectric shifters are constrained by frequency response, particularly for short coils \cite{trommer1990passive,trommer1996passive}. These challenges are not unique to IFOGs but apply to many optical gyroscopes relying on active biasing. Therefore, a passive, robust, and low-maintenance phase biasing approach is needed to eliminate these negative effects of active modulation.

Various passive biasing techniques have been proposed over the decades to address the challenges of active modulation in interferometric fiber optic gyroscopes (IFOGs). Early efforts, such as Sheem’s 1980 use of a 3×3 directional coupler \cite{sheem1980fiber}, eliminated active modulators but introduced complexity and sensitivity issues related to asymmetry and signal loss. Subsequent developments by Kajioka and Matsumura in 1984 utilized polarization state changes to passively detect rotation-induced phase shifts, though environmental susceptibility remained a limitation \cite{kajioka1984single}. Other approaches, including those by Huang in 2007 and Jabo and Xiao Qian in 2009, implemented specialized fiber couplers and branching units for passive biasing but faced challenges with fabrication complexity and tuning precision \cite{huang2010passively,Bo2009}. More recently, Hakimi et al. in 2004 advanced the concept with a topological phase bias element, yet issues such as polarization mode dispersion and drift limited practical performance \cite{hakimi2024passive}.

Despite these efforts, existing passive biasing methods have not fully achieved the combination of high sensitivity, robustness, and simplicity required for reliable phase biasing near $\pi/2$ in IFOGs. Notably, Steve Yao’s in 2016 design demonstrated an energy-efficient passive IFOG system employing a polarization beam splitter and retarder to realize a stable $\pi/2$ phase bias, laying important groundwork \cite{yao2016energy}.

Building on this foundation, our latest paper demonstrated, for the first time based on our knowledge, a passive-biased IFOG enabling simultaneous operation at two quadrature points ($\pi/2$ and $3\pi/2$) with inherent noise suppression using a Non-Reciprocal Polarization Dependent Phase Shifter (NRPPS) \cite{akcaalan2025phasebiasingopticalgyroscope}. The present work extends this approach by introducing double sensitivity operation while maintaining full passive functionality for the first time based on our knowledge. Through detailed theoretical analysis we show that the enhanced double sensitive NRPPS-IFOG which is called DS-NRPPS-IFOG outperforms conventional designs and also double sensitive designs in terms of sensitivity, stability, and noise resilience. Additionally, recent studies have demonstrated that the Shupe effect \cite{cao2023dual} and magnetic effects \cite{liu2017drift,yertutanol2021fiber} can be mitigated by simultaneously employing two orthogonal polarizations as horizontal and vertical, in the sensing coil. Thus, the DS-NRPPS-IFOG naturally lends itself as a strong candidate for achieving these suppression. These advancements represent a key step toward compact, low-power, and high-precision passive-biased gyroscopes for demanding navigation applications.

\begin{figure}[!b]
\centering
\includegraphics[width=0.8\textwidth]{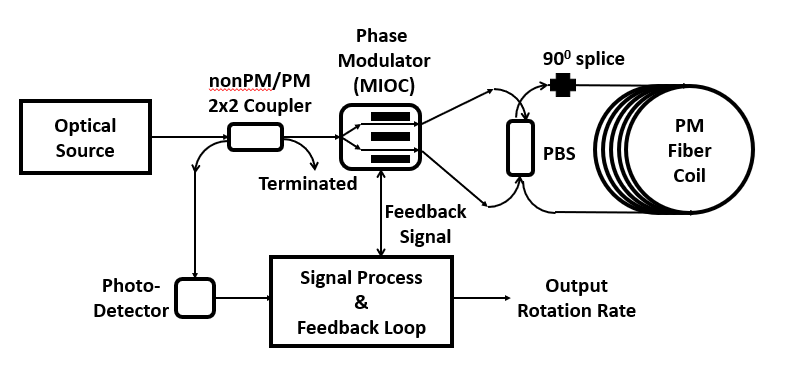}
\caption{\label{fig:DS-IFOG_Setup} The minimum configuration of DS-IFOG setup}
\end{figure}

In the first section of the paper offers a clear comparison between the first double sensitivite IFOG configuration, called DS-IFOG \cite{zhou2013fiber} which is also pioneer for the other double sensitive IFOGs \cite{wu2018open,zhang2020novel}, and the proposed DS-NRPPS-IFOG system, by emphasizing the primary differences in their architectures. The second section provides an in-depth explanation of the NRPPS in DS-NRPPS-IFOG system and how it passively introduces a phase bias between the clockwise (CW) and counter-clockwise (CCW) light beams within the IFOG. Lastly, we evaluate the DS-NRPPS-IFOG’s performance through simulations, comparion it against a traditional DS-IFOG with respect to sensitivity, accuracy, and long-term stability. The simulation outcomes validate the proposed design’s effectiveness, showcasing its potential for improved stability, enhanced sensitivity, and noise reduction in real-world applications.

In the minimum configuration of DS-IFOG \cite{zhou2013fiber} consists of a coherent light source, a beam splitting and recombination unit, a multifunctional integrated optical chip (MIOC - served as a 2×2 polarization-maintaining (PM) coupler, a polarizer, and a modulator), a polarization splitter/combiner, a sensing coil, and a photodetector system, all connected through polarization-maintaining (PM) fiber as seen in Fig-~\ref{fig:DS-IFOG_Setup}.

Light from a coherent source is first launched into a fiber optic coupler or splitter and then directed to a MIOC, which splits the beam into two paths traveling in opposite directions. Instead of reaching the sensing coil directly, these beams are routed to a 2×2 fiber coupled polarization beam splitter (PBS). The outputs of the PBS are connected to the polarization-maintaining (PM) sensing coil; however, one of the arms includes a 90-degree splice that rotates the beam’s polarization orientation. This deliberate polarization rotation enables the beams to pass through the sensing coil a second time, creating a double-sensitive section that enhances the system’s overall sensitivity. To ensure a linear system response near zero rotation, a phase bias of $\pi/2$ is introduced between the clockwise (CW) and counter-clockwise (CCW) beams via MIOC component. 

\begin{figure}[!t]
\centering
\includegraphics[width=0.95\textwidth]{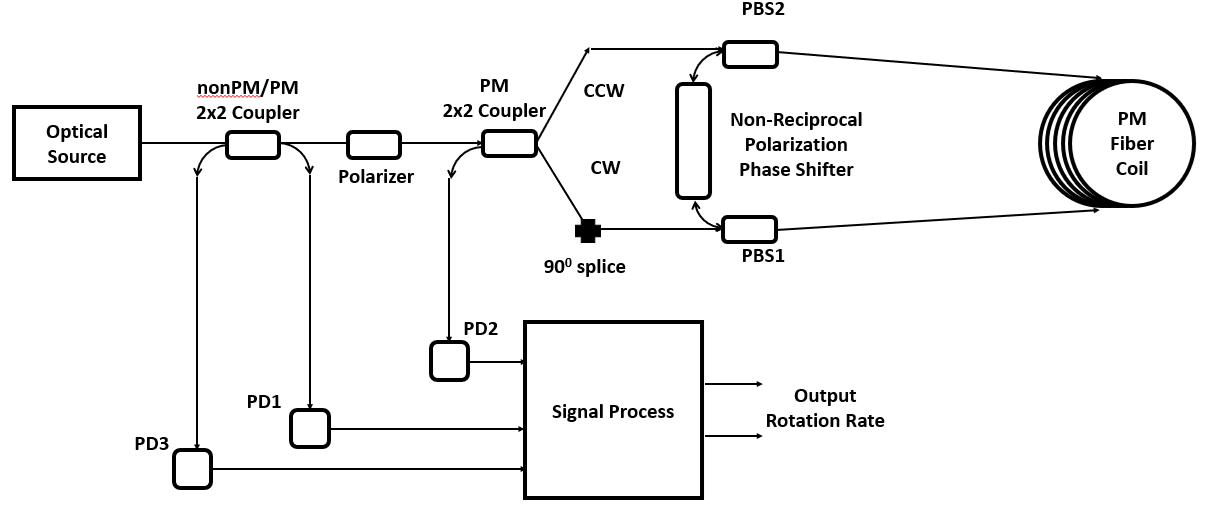}
\caption{\label{fig:DS-NRPPS-IFOG_Setup} A Non-Reciprocal Polarization Phase Shifter based DS-IFOG system called DS-NRPPS-IFOG.}
\end{figure}

As illustrated in Fig.~\ref{fig:DS-NRPPS-IFOG_Setup}, the DS-NRPPS-based IFOG system employs a passive, non-reciprocal polarization phase-shifting technique to substitute traditional active modulation components. The DS-NRPPS-IFOG configuration begins with a coherent light source linked to an optional non-PM or PM 2×2 coupler, selected based on whether the source is linearly polarized. A polarizer is used to ensure the beam is linearly polarized before it enters the main PM 2×2 coupler, which divides the light into clockwise (CW) and counterclockwise (CCW) propagation paths. 

\begin{figure}[!b]
\centering
\includegraphics[width=0.8\textwidth]{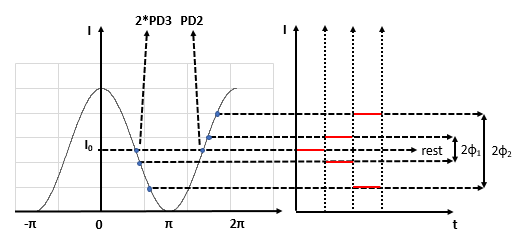}
\caption{\label{fig:Sinus_Response} Output signals of PD2 and PD3 section for DS-NRPPS-IFOG which is similar to NRPPS-IFOG, while rest and with rotation $\phi_{1}$ and $\phi_{2}$ \cite{akcaalan2025phasebiasingopticalgyroscope}.}
\end{figure}

The CW beam is directed to the vertical-polarization port of the first fiber-coupled PBS via a $90^\circ$ rotated splice located between PM 2×2 coupler and the first PBS. By passing the vertical polarization port of the PBS (PBS1), the beam then passes through the PM fiber coil and reaches the second fiber-coupled PBS (PBS2) through its vertical polarization port. Between PBS1 and PBS2 lies a Non-Reciprocal Polarization Phase Shifter (NRPPS), responsible for both introducing a polarization-dependent phase shift and the second round for the PM fiber coil. After passing through NRPPS, the beam’s polarization is rotated to horizontal and it returns to the horizontal polarization port of PBS1. The beam subsequently traverses PM fiber coil and PBS2 again before returning to PM 2×2 coupler.

The CCW beam, emerging from PM 2×2 coupler, travels through the horizontal port of PBS2, then follows the path through PM fiber coil and PBS1 sequentially. Since the horizontal port of PBS1 connects to the NRPPS. This beam also passes through NRPPS, undergoing a polarization rotation to vertical. It then returns to the vertical polarization port of PBS2 and continues through PM fiber coil and PBS1. Before reaching back to PM 2×2 coupler, the $90^\circ$ rotated splice rotates the CCW beam polarization from vertical to horizontal and ensures that the CCW beam reaches to PM 2×2 coupler with horizontal polarization.

After undergoing phase shifts in NRPPS, CW and CCW beams recombine at PM 2×2 coupler with proper polarization alignment to produce interference signals detected by photodetectors as PD2 and PD3. The first system output at coupler PD2 contains a phase shift composed of both the $\pi$ phase shift introduced by PM 2×2 coupler and the additional polarization-dependent phase shift, $\phi_r$ from NRPPS. The combined signal at PD2 represents the sum of the CW and CCW amplitudes, further modulated by the Sagnac phase shift $\phi_{i}$:

\begin{eqnarray}
\centering
&&\label{eqnarray:1} E_{PD2} = E_{CW}e^{0j} + E_{CW}e^{-j(\phi_i+(\pi+\phi_{r}))},\\
assume
&&\label{eqnarray:2} E_{CW}=E_{CCW}=E,\\
&&\label{eqnarray:3}E_{PD2}=E(\cos{(0^\circ)}+j\sin{(0^\circ)}+\cos{(\phi_{i}+(\pi + \phi_{r}))}-j\sin{(\phi_{i}+(\pi + \phi_{r}))}).
\end{eqnarray}

The intensity on the PD2 detector becomes;

\begin{eqnarray}
\centering
&&\label{eqnarray:4} I_{PD2} = |E_{PD2}|^2 = 2I(1+\cos{(\phi_{i}+(\pi + \phi_{r}))}).
\end{eqnarray}

An additional measurement is taken from the output of the optional coupler as PD3, which carries only the phase shift, $\phi_r$ introduced by NRPPS, excluding the $\pi$ phase difference from coupler PM 2×2 coupler. The sum of the CW and CCW amplitudes, shifted by the Sagnac phase $\phi_{i}$, is detected at PD3:

\begin{eqnarray}
\centering
&&\label{eqnarray:5} E_{PD3} = E_{CW}e^{0j} + E_{CW}e^{-j(\phi_i+\phi_{r})},\\
assume
&&\label{eqnarray:6} E_{CW}=E_{CCW}=E,\\
&&\label{eqnarray:7} E_{PD3}=E(\cos{(0^\circ)}+j\sin{(0^\circ)}+\cos{(\phi_{i}+ \phi_{r})}-j\sin{(\phi_{i}+\phi_{r}})).
\end{eqnarray}

The intensity on the PD3 detector becomes;

\begin{eqnarray}
\centering
&&\label{eqnarray:8} I_{PD3} = |E_{PD3}|^2 = 2I(1+\cos{(\phi_{i}+ \phi_{r})}).
\end{eqnarray}

This configuration facilitates two independent rotation measurements at different quadrature points. These measurements can be used individually or combined to achieve enhanced sensitivity. In the case of usage both photo-detectors and When the retarder is configured as a quarter-wave plate at $45^\circ$, which corresponds to $\phi_{r}=\pi/2$ the system produces output signals that are in quadrature ($\pi/2$ and $3\pi/2$, as shown in Fig.\ref{fig:Sinus_Response}). This ensures a linear sinusoidal response, complementary detection, and improved sensitivity—all achieved without active modulation. The retarder can be adjusted to other phase values based on user requirements, but setting it to $\pi/2$ provides a same phase shift of $\pi/2$. Due to their different positions in the system, the intensities at PD2 and PD3 differ; neglecting component losses, the intensity at PD3 is expected to be half that of PD2, as illustrated in Fig.\ref{fig:Sinus_Response}. Using quadrature of $\pi/2$ and $3\pi/2$, is shown for the first time in NRPPS-IFOG paper, which also allows noise cancellation by adjusting the temporal offset between the signals from PD2 and PD3 \cite{akcaalan2025phasebiasingopticalgyroscope}.

\begin{figure}[!t]
\centering
\includegraphics[width=0.95\textwidth]{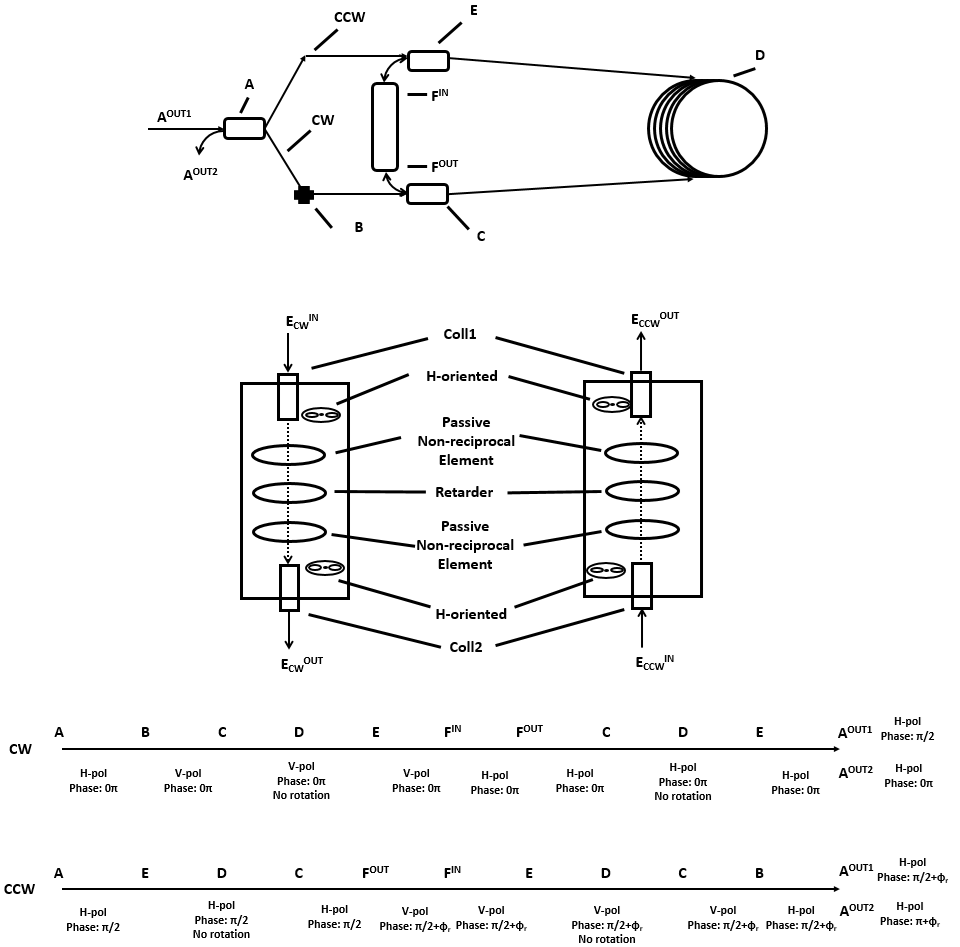}
\caption{\label{fig:CW_CCW-DS-NRPPS} CW and CCW beam paths for Double Sensitive Non-Reciprocal Polarization Phase Shifter Design}
\end{figure}

From this point forward, the analysis will focus on an DS-NRPPS configuration that incorporates an quarter-wave plate as a retarder, oriented such that the polarization of the CW beam aligns with the retarder’s fast axis. The analysis will begin by defining the Jones matrices for the optical components in terms of polarization and phase. Then, the behaviors of the CW and CCW beams will be derived based on these definitions, according to the DS-NRPPS design.

\begin{eqnarray}
\centering
\textbf{horizontal polarization}  
&&\label{eqnarray:9} 
  H =
  \left[ {\begin{array}{cc}
    1 \\
    0 \\
  \end{array} } \right],\\
  \textbf{vertical polarization}
&&\label{eqnarray:10} 
  V =
  \left[ {\begin{array}{cc}
    0 \\
    1 \\
  \end{array} } \right],\\
\textbf{PBS}
&&\label{eqnarray:11} 
  J_{PBS} =
  \left[ {\begin{array}{cc}
    1 & 0\\
    0 & 1\\
  \end{array} } \right],\\
  \textbf{Rotation matrix}
&&\label{eqnarray:12} 
  R(\theta) =
  \left[ {\begin{array}{cc}
    cos(\theta) & -sin(\theta)\\
    sin(\theta) & cos(\theta)\\
  \end{array} } \right],\\
    \textbf{Collimator rotation}
&&\label{eqnarray:13} 
  J_{coll}(\theta) =
  \left[ {\begin{array}{cc}
    cos(\theta) & -sin(\theta)\\
    sin(\theta) & cos(\theta)\\
  \end{array} } \right],\\
 \textbf{Passive Non-Reciprocal Element at $R(45^o )$}
&&\label{eqnarray:14} 
  J_{non-reciprocal El.} = 1/\sqrt{2}
  \left[ {\begin{array}{cc}
    1 & -1\\
    1 & 1\\
  \end{array} } \right],\\
\textbf{Retarder at $R(R_1)$}
&&\label{eqnarray:15} 
  J_{retarder} = R(-R_1 )
  \left[ {\begin{array}{cc}
    e^{i\phi_r} & 0\\
    0 & e^{-i\phi_r}\\
  \end{array} } \right]R(R_1 ),\\
    \textbf{Splice - a projection matrix}
&&\label{eqnarray:16} 
  J_{splice} =
  \left[ {\begin{array}{cc}
    cos^2(\theta) & -cos(\theta)sin(\theta)\\
    sin(\theta)cos(\theta) & sin^2(\theta)\\
  \end{array} } \right].
\end{eqnarray}

Fig-~\ref{fig:CW_CCW-DS-NRPPS} illustrates how the Non-Reciprocal Polarization Phase Shifter Design is working with respect to CW and CCW beam for the DS-NRPPS design. For the CW beam which comes from one port of the 2x2 PM coupler with horizontal polarization and lets say $0\pi$ relative phase shift, follows the CW beam path as shown in Fig-~\ref{fig:CW_CCW-DS-NRPPS}. Here is the Jones matrix of CW beam according to the DS-NRPPS design:

\begin{eqnarray}
\centering
&&\label{eqnarray:17} 
  E_{in}^{CW}=H =
  \left[ {\begin{array}{cc}
    1 \\
    0 \\
  \end{array} } \right],\\
  &&\label{eqnarray:18} 
 J_{PBS}*J_{PBS}*J_{coll2}(0^\circ)*J_{non-reciprocal El.}(45^\circ)*J_{retarder}(R_1)*  \nonumber\\
 && *J_{non-reciprocal El.}(45^\circ)*J_{coll1}(0^\circ)*J_{PBS}* J_{PBS}*J_{splice}(90^\circ)*E_{in}^{CW} =E_{out}^{CW},\\
 &&\label{eqnarray:19} 
  E_{out}^{CW}=
  \left[ {\begin{array}{cc}
    1 \\
    0 \\
  \end{array} } \right].
\end{eqnarray}

\begin{table}[!b]
\centering
\caption{(left) Performance comparison of conventional DS-IFOG and DS-NRPPS-IFOG in terms of rotation rate and (right) Overlapping Allan variance analysis for 200m., 1000m. and, 2000m. fiber spool.}
\label{tab:table1}
\begin{tabular}{|c|c|c|}
\hline
\parbox{3cm}{$200m.$ \\
\\
ARW (DS-IFOG):  \\
$0.00380 ^\circ/\sqrt{hr}$ \\
\\
ARW (DS-NRPPS): \\
$0.00025 ^\circ/\sqrt{hr}$ \\} 
  & \raisebox{-0.5\height}{\includegraphics[height=4cm]{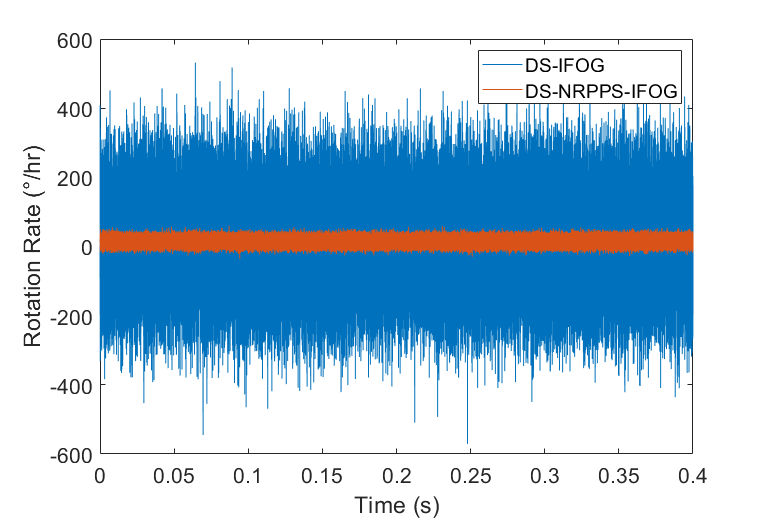}} 
  & \raisebox{-0.5\height}{\includegraphics[height=4cm]{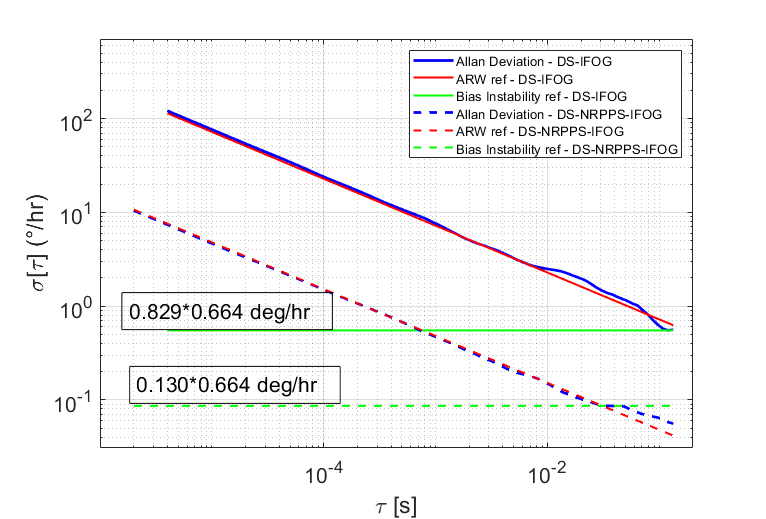}} \\
  \parbox {3cm}{$1000m.$ \\
  \\
ARW (DS-IFOG):  \\
$0.00081 ^\circ/\sqrt{hr}$ \\
\\
ARW (DS-NRPPS): \\
$0.00002 ^\circ/\sqrt{hr}$ \\} 
  & \raisebox{-0.5\height}{\includegraphics[height=4cm]{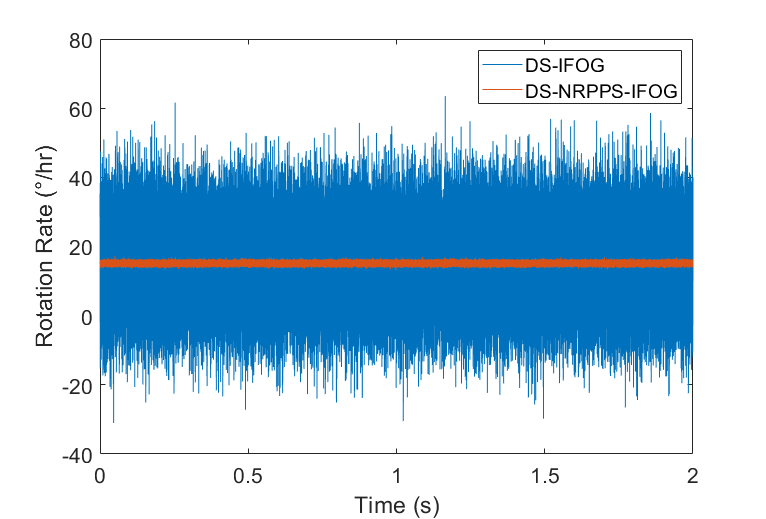}} 
  & \raisebox{-0.5\height}{\includegraphics[height=4cm]{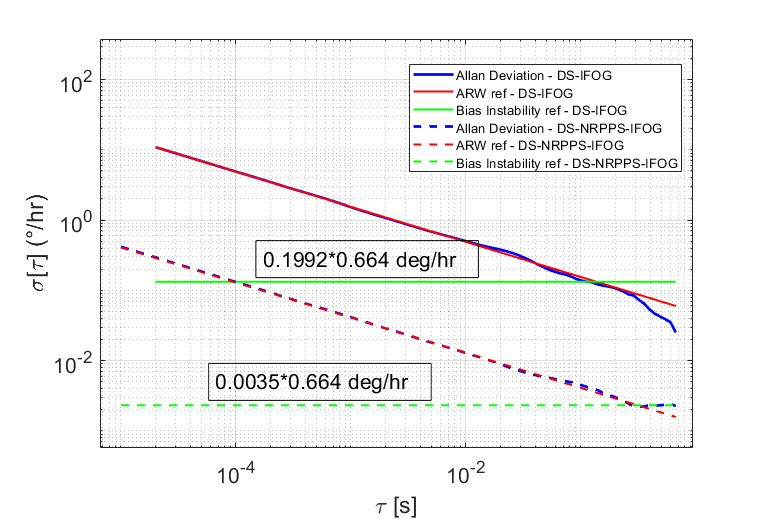}} \\
    \parbox {3cm}{$2000m.$ \\
    \\
ARW (DS-IFOG):  \\
$0.000410 ^\circ/\sqrt{hr}$ \\
\\
ARW (DS-NRPPS): \\
$0.000008 ^\circ/\sqrt{hr}$ \\} 
  & \raisebox{-0.5\height}{\includegraphics[height=4cm]{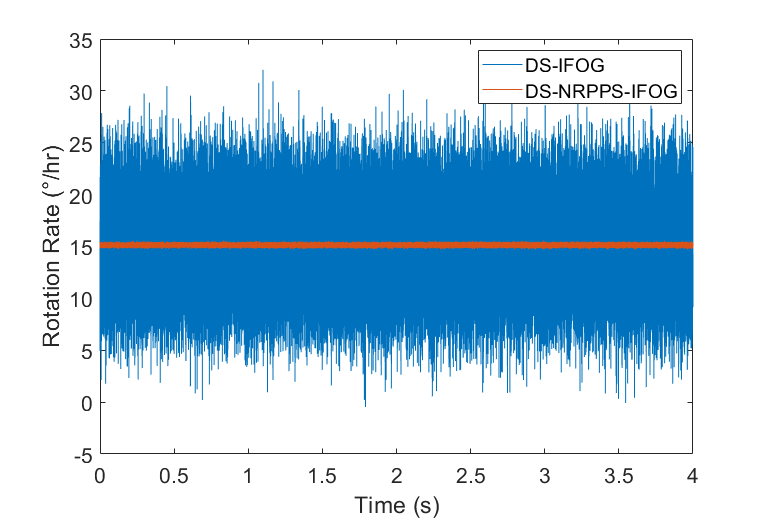}} 
  & \raisebox{-0.5\height}{\includegraphics[height=4cm]{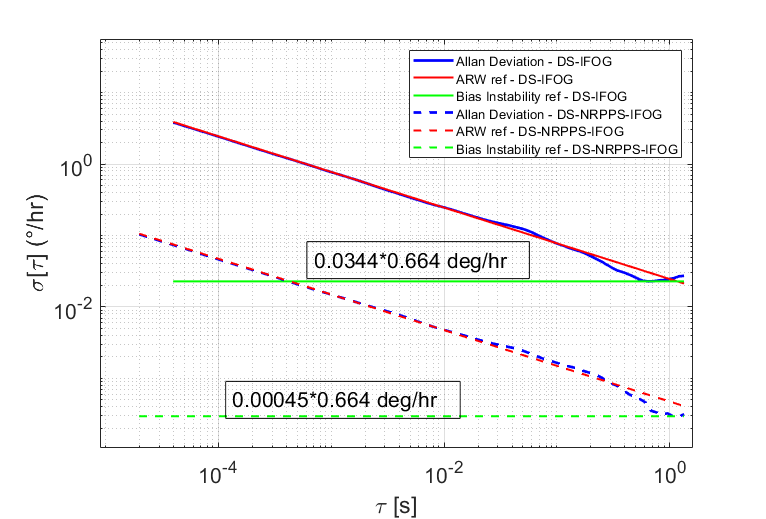}} \\
\hline
\end{tabular}
\end{table}

For the CCW beam which comes from one port of the 2x2 PM coupler with horizontal polarization and a $\pi/2$ relative phase shift as a result of evanescent coupling mechanism, follows the CCW beam path as shown in Fig-~\ref{fig:CW_CCW-DS-NRPPS}. Here is the Jones matrix of CCW beam according to the DS-NRPPS design:

\begin{eqnarray}
\centering
&&\label{eqnarray:17} 
  E_{in}^{CCW}=e^{j\pi/2}H =e^{j\pi/2}
  \left[ {\begin{array}{cc}
    1 \\
    0 \\
  \end{array} } \right],\\
  &&\label{eqnarray:18} 
 J_{splice}(90^\circ)*J_{PBS}*J_{PBS}*J_{coll1}(0^\circ)*J_{non-reciprocal El.}(45^\circ)* \nonumber\\
 && J_{retarder}(R_1)*J_{non-reciprocal El.}(45^\circ)*J_{coll2}(0^\circ)*J_{PBS}* J_{PBS}*E_{in}^{CCW} =E_{out}^{CCW},\\
 &&\label{eqnarray:19} 
  E_{out}^{CCW}= e^{j(\pi/2+\phi_r)}
  \left[ {\begin{array}{cc}
    1 \\
    0 \\
  \end{array} } \right].
\end{eqnarray}

In equations (4) and (8), the interfered signals on PD2 and PD3 are already shown. As a result, the DS-NRPPS introduces a both double sensitive and a precise and absolute $\pi/2$ phase shift between $E_{out}^{CW}$  and $E_{out}^{CCW}$, effectively shifting the operating point into the linear region of the sinusoidal response.

The simulations were conducted using a custom MATLAB code, which was also utilized in the previous study on NRPPS-IFOG and the details for the optical source is already given in NRPPS-IFOG paper \cite{akcaalan2025phasebiasingopticalgyroscope}. Besides the components used in a conventional IFOG setup, for the minimum configuration of the DS-IFOG,  a PBS and includes a $90^\circ$ rotated splice are added to reroute the CW and CCW beams for a second turn through the fiber coil, as illustrated in Fig-~\ref{fig:DS-IFOG_Setup}.

For the DS-NRPPS-IFOG, the same optical source was utilized. A polarizer with a polarization extinction ratio (PER) of 30 dB was placed ahead of the 2×2 PM coupler connected to the sensing coil. In the NRPPS section, an quarter-wave plate as a retarder was used to achieve a $\pi/2$ and $3\pi/2$ phase shifts at the outputs. As shown in the system diagram, there are three output ports: two for extracting rotation-induced phase shifts and one for monitoring the input beam. Photodetectors, PD2 and PD3 (as indicated in Fig.-~\ref{fig:DS-NRPPS-IFOG_Setup}) were employed to measure the rotation rate, with a deliberate 3-meter fiber length difference introduced to create a temporal offset between the signals as mentioned NRPPS-IFOG as well \cite{akcaalan2025phasebiasingopticalgyroscope}. The another key advantage of the DS-NRPPS-IFOG configuration over the DS-IFOG lies in its capability for continuous rotation measurement, allowing the system to acquire one data point per modulation cycle after averaging over the entire cycle.

\begin{table}[!b]
\centering
\caption{Performance comparison of the DS-NRPPS-IFOG in terms of temporal offset between the signals of PD3 and PD2 for 2000m. fiber coil.}
\label{tab:table2}
\begin{tabular}{|c|c|c|}
\hline
\parbox{3cm}{Rotation Rates} 
  & \raisebox{-0.5\height}{\includegraphics[height=6cm]{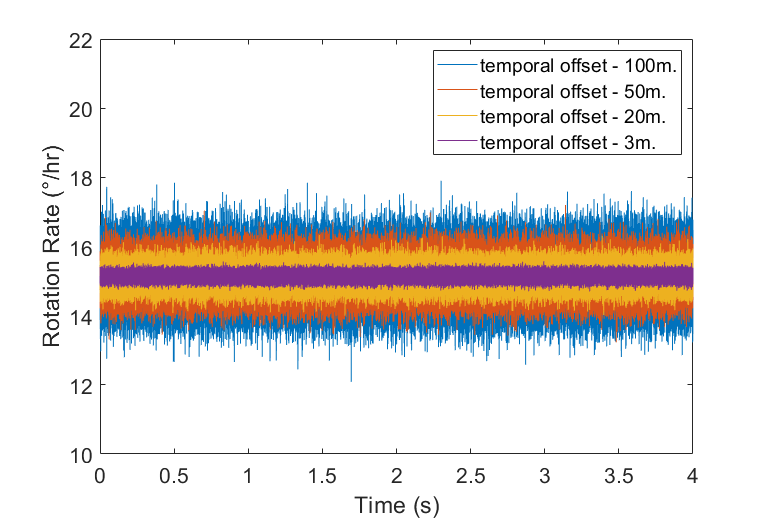}} \\
  \parbox {3cm}{$ARW - 3m.$ \\
$0.000008 ^\circ/\sqrt{hr}$ \\
\\
$ARW - 20m.$ \\
$0.000020 ^\circ/\sqrt{hr}$ \\
\\
$ARW - 50m.$ \\
$0.000032 ^\circ/\sqrt{hr}$ \\
\\
$ARW - 100m.$ \\
$0.000045 ^\circ/\sqrt{hr}$ \\}
  & \raisebox{-0.5\height}{\includegraphics[height=6cm]{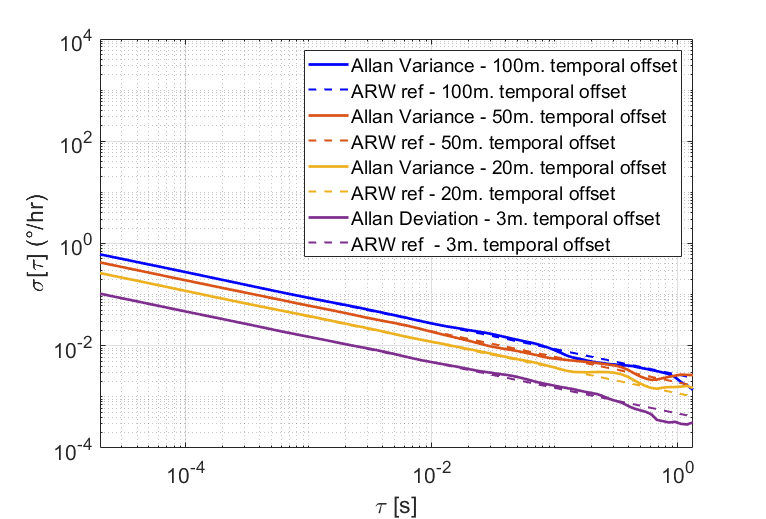}} \\
\hline
\end{tabular}
\end{table}

The simulation have sensitivity measurements based on 200,000 round trips through the detection coil, assuming a constant rotation rate equivalent to the Earth's rotation as $15^\circ/s$. For each cycle along the coil, a single data sample was captured at approximately $1/c$ intervals, where $c$ represents the speed of light. This sampling interval was kept consistent across all system configurations. In the case of the conventional DS-IFOG, one averaged data point was obtained over two full coil cycles (four cycles of the fiber coil due to double pass). In contrast, the DS-NRPPS-IFOG configuration enabled one data point to be captured per full coil cycle (double cycles due to double pass), owing to its ability to extract a measurement from each pass. To assess the impact of polarization-maintaining (PM) coil length on system performance, simulations were performed using coil lengths of 200 m, 1000 m, and 2000 m, with all fiber coils wound to a diameter of 10 cm, as summarized in Table~\ref{tab:table1}. Additionally, for comparison of the results between the conventional IFOG and DS versions, one can check it from our previous paper \cite{akcaalan2025phasebiasingopticalgyroscope}.

As a result, the simulation shows that the Angular Random Walk (ARW) values for the DS-NRPPS-IFOG are approximately 15x, 40x, and 50x lower than those of the conventional DS-IFOG for fiber lengths of 200 m, 1000 m, and 2000 m, respectively. This marked enhancement in sensitivity is largely due to the utilization of PD2 and PD3, which detect the same signal with a relative phase difference of $\pi$, effectively capturing quadrature components of the sinusoidal response. Furthermore, the ability to fine-tune the temporal offset between the outputs of PD2 and PD3 allows for improved noise suppression. Our simulation results highlight that the system’s overall performance is highly sensitive to this temporal offset, as detailed in Table~\ref{tab:table2}.

\section{Conclusion}
In conclusion, the DS-NRPPS-IFOG introduces, for the first time to the best of our knowledge, a both double sensitive and fully passive $\pi/2$ phase modulation scheme capable of operating simultaneously at two quadrature points: $\pi/2$ and $3\pi/2$. This architecture has been analytically evaluated and compared with a conventional DS-IFOG configuration. The results demonstrate a substantial improvement in performance, with Angular Random Walk (ARW) values up to 50× lower than those of conventional DS-IFOGs, depending on the fiber coil length. While this study focuses on comparing DS-NRPPS-IFOG with traditional DS-IFOG systems, the proposed non-reciprocal passive biasing mechanism shows promising applicability in other types of optical gyroscopes that currently rely on active modulation. Additionally, the ability to operate simultaneously at dual quadrature points inherently enables spontaneous noise cancellation, further enhancing measurement stability and sensitivity.

\section*{Acknowledgments}
A provisional patent application has been filed for the technology described in this paper. O.A. conducted the simulations, analyzed the results and supervised the manuscript. M.G.A. reviewed the manuscript.

\bibliographystyle{unsrt}
\bibliography{DS-NRPPS-IFOG}

\end{document}